\documentclass[fleqn]{article}
\DeclareFontFamily{OT1}{rsfs}{}
\DeclareFontShape{OT1}{rsfs}{m}{n}{<5> rsfs5 <7> rsfs7 <10> rsfs10
}{}
\DeclareSymbolFont{mathrsfs}{OT1}{rsfs}{m}{n}
\DeclareSymbolFontAlphabet{\mathrsfs}{mathrsfs}
\begin{document}

\author{R.R. Lompay\thanks{Institute of Electron Physics Nat.Acad.Sci.of Ukraine,
Department of Theory of Elementary Interactions, Uzhgorod,
Ukraine,  email: lompay@zk.arbitr.gov.ua} .}
\title{Propertime theories and reparametrization-invariant theories}
\date{}
\maketitle
\newcommand{\Lag}{\mathrsfs{L}}
\newcommand{\Ham}{\mathrsfs{H}}
\newcommand{\xz}{\delta(x-z)}
\newcommand{\sym}[1]{\stackrel{sym}{#1}{}}
\newcommand{\dt}{\mathrsfs{D}\theta\,}
\newcommand{\limz}{_{\theta_1}^{\theta_2}}
\newcommand{\limx}{_{\Sigma_1}^{\Sigma_2}}
\newcommand{\limi}{_{-\infty}^{+\infty}}
\newcommand{\intz}{\int\limits\limz\dt }
\newcommand{\intx}{\int\limits\limx dx\,}
\newcommand{\ints}{\int\limits_\Sigma d\sigma_\mu\,}
\newcommand{\intzi}{\int\limits\limi\dt }
\newcommand{\imp}[2]{\frac{\partial L^I}{\partial Dz^#1}(\delta^#1{}_#2+Dz^#1 Dz_#2)-L^I Dz_#2}
\begin{abstract}
Reparametrization-invariant theories of point relativistic
particle interaction with fields of arbitrary tensor dimension
are considered. It has been shown that the equations of motion
obtained by Kalman [G. Kalman, Phys. Rev. {\bf 123}, 384 (1961)]
are reproduced as the Euler-Lagrange equations for
reparametrization- invariant theory in the propertime gauge. The
formalism is developed that conserves manifest reparametrization
invariance at each stage of calculations. Using the above
formalism, the equations of motion are being analyzed and the
dynamical variable theories are being constructed. It has been
shown that the remained invariance kept after gauge fixation
gives an identically-zero invariant Hamiltonian.

PACS number(s): 11.30.-j, 11.10.Ef, 03.30.+p, 03.50.-z
\end{abstract}
\section*{Introduction}

The task of obtaining the equations of motion for a point
relativistic particle that moves in the external field of
specified tensor dimensionality was analyzed by Kalman
\cite{Kal}. On the strength of the covariant Lagrange formalism,
the equations of motion of the particle were derived in his
approach both by varying the trajectory of the particle and by
manifest varying its propertime. This approach agrees well with
geometric interpretation in the space-time, but equations of
motion obtained by Kalman differ from Euler - Lagrange equations
by the terms resulted from the manifest variation of the particle
propertime. In this paper, the theories of a point relativistic
particle interaction with arbitrary tensor dimensionality field
are considered having reparametrization invariance (RI). In such
theories, the action functional is not changed at arbitrary
functional replacement of the parameter that numerates the points
of the particle world line. It appears that the equations of
motion of the particle obtained by varying only the particle
trajectory with no variation of its time parameter (i.e. relevant
Euler-Lagrange equations) coincide with those obtained by Kalman,
if one sets in the end of calculations the parameter equal to the
propertime of the particle. Thus, the Kalman equations could be
reproduced as Euler-Lagrange equations in the
reparametrization-invariant theory in the propertime gauge. This
paper consists of two parts. The first part is dedicated to the
obtaining and analysis of equations of motion for the
relativistic particle that moves in the external field of
specified tensor dimensionality. Paragraph 1.1 deals with a brief
derivation of the Kalman equations. In paragraph 1.2, the general
form of RI-action functional of the particle is considered, Euler
- Lagrange equations are derived in the RI-theories and their
equivalence with the Kalman equations in the propertime gauge is
demonstrated. The formalism that conserves the manifest
reparametrization invariance at each stage of calculations is
developed. Paragraph 1.3 shows that N\H{o}ether's equality
corresponding to the RI theory results in the presence in the
equations of motion of the particle of the projector onto its
proper 3-space. The notions of dynamical (effective) mass of the
particle and the 4-force acting on it are introduced.

The second part deals with the construction of the theory of a
total dynamical system consisting of the interacting point
relativistic particle and of the field of arbitrary tensor
dimensionality. The above system has an manifest RI. In paragraph
2.2, the total variation of the action functional for the system
is found, the equations of motion of the particle and field are
derived. The system 4-momentum vector and the canonical
energy-momentum tensor (EMT) are constructed in the paragraph
2.2. The conservative character of canonical EMT is proven. In
paragraph 2.3, the 4-angular momentum tensor and the relevant
density tensor are constructed. The symmetric EMT is constructed
in paragraph 2.4 by generalizing the Belinfante method \cite{Bel}
for the case of the systems consisting of the interacting point
particle and fields. It has been shown in the paragraph 2.5 that
the remained invariance kept after gauge fixation and related to
the freedom of choice of an initial value of the parameter gives
the zero N\H{o}ether's current, and, hence, does not lead to the
integral of motion that is additional to the 10 relativistic
ones. All the considerations presented in this part have been
carried out by Lorentz-invariant and RI-manner.

The following notations are used in this paper. Greek indices
$\alpha,\beta,\dots,\mu,\nu,\dots$ run over the values from 0 to
3. The $(-1,1,1,1)$ metric signature is chosen. The velocity of
light in vacuum is set equal to unit: $c=1$.
\section{Equations of motion of the particles}
\setcounter{equation}{0}
\subsection{Kalman equations}

It seems natural to assume that the action functional for the
particle that moves in the external fields has a form:
\begin{equation}
\begin{array}{l}
W[z;\phi]=W^P[z]+W^I[z;\phi]=\int\limits_{z_1}^{z_2}d\tau\,L^P(z(\tau),\dot{z}(\tau))\\
 \qquad+\intx\int\limits_{z_1}^{z_2}d\tau\,L^I(z(\tau),\dot{z}(\tau);\phi(x),\partial\phi(x))
\delta(x-z(\tau)),
\end{array}
\end{equation}
where $\tau$ is the propertime of the particle multiplied by $c$.
$z\equiv\{z^\mu\}\equiv(z^0,z^1,z^2,z^3)$ are pseudo-cartesian
coordinates of the particle in the Minkowski space;
$\dot{z}\equiv\{\dot{z}^\mu\}\equiv\{dz^\mu(\tau)/d\tau\}$;
$\phi$ is a set of external fields; $dx\equiv dx^0 dx^1 dx^2
dx^3$,
$\delta(x)\equiv\delta(x^0)\delta(x^1)\delta(x^2)\delta(x^3)$.
Let us denote
\begin{equation}
L\equiv L^P(z,\dot{z})+\intx
L^I(z,\dot{z};\phi(x),\partial\phi(x))\xz\equiv L(z,\dot{z}).
\end{equation}
Let us derive a generalized equation of motion of the point
particle using variational principle. Since $\tau$ is an
propertime of the particle, then variation of its trajectory also
causes relevant variation of the propertime "element". Therefore,
assuming boundary points to be non-variable, we find
\begin{equation}
\delta W=\delta\int\limits_{z_1}^{z_2}d\tau\,L=
\int\limits_{z_1}^{z_2}\delta d\tau\,L+
\int\limits_{z_1}^{z_2}d\tau\,\delta L,
\end{equation}
where
\begin{equation}
\delta L=\frac{\partial L}{\partial z^\mu}\delta z^\mu+
\frac{\partial L}{\partial \dot{z}^\mu}\delta \dot{z}^\mu.
\end{equation}
For $\delta\dot{z}^\mu$
\begin{equation}
\delta\dot{z}^\mu=\delta\left(\frac{dz^\mu}{d\tau}\right)=
\frac{\delta dz^\mu}{d\tau}-\frac{dz^\mu}{(d\tau)^2}\delta d\tau.
\end{equation}
Since $\tau$ is an propertime of the particle, the following
relation holds true
\begin{equation}
d\tau=\sqrt{-dz^\alpha dz_\alpha},
\end{equation}
where we find
\begin{equation}
\delta d\tau=-\frac{dz_\mu}{\sqrt{-dz^\alpha dz_\alpha}}\delta
dz^\mu= -\dot{z}_\mu\delta dz^\mu.
\end{equation}
Substituting (1.7) into (1.5), we obtain
\begin{equation}
\delta\dot{z}^\mu=\frac{\delta dz^\mu}{d\tau}+
\dot{z}^\mu\dot{z}_\nu\frac{\delta dz^\nu}{d\tau}=
(\delta^\mu{}_\nu+\dot{z}^\mu\dot{z}_\nu)\frac{\delta
dz^\nu}{d\tau}.
\end{equation}
However, $\delta dz^\mu=d\delta z^\mu$, then
\begin{equation}
\frac{\delta dz^\mu}{d\tau}=\frac{d}{d\tau}\delta z^\mu.
\end{equation}
Therefore
\begin{equation} \delta d\tau=-\dot{z}_\mu d\delta z^\mu.
\end{equation}
Thus, according to (1.8), (1.10) and (1.4)
\begin{equation}
\delta\dot{z}^\mu=(\delta^\mu{}_\nu+\dot{z}^\mu\dot{z}_\nu)\frac{d}{d\tau}\delta
z^\nu,
\end{equation}
\begin{equation}
\delta L=\frac{\partial L}{\partial z^\mu}\delta z^\mu+
\frac{\partial L}{\partial \dot{z}^\mu}
(\delta^\mu{}_\nu+\dot{z}^\mu\dot{z}_\nu)\frac{d}{d\tau}\delta
z^\nu.
\end{equation}
Substituting now (1.10) and (1.12) into (1.3) and integrating by
parts, we find
\begin{equation}
\begin{array}{l}
\delta W=- \int\limits_{z_1}^{z_2}L\dot{z}_\mu d\delta z^\mu+
\int\limits_{z_1}^{z_2}d\tau\,\left[\frac{\partial L}{\partial
z^\mu}\delta z^\mu+\frac{\partial L}{\partial \dot{z}^\mu}
(\delta^\mu{}_\nu+
\dot{z}^\mu\dot{z}_\nu)\frac{d}{d\tau}\delta z^\nu\right]\\
\qquad=\left.\left[\frac{\partial L}{\partial \dot{z}^\mu}
(\delta^\mu{}_\nu+\dot{z}^\mu\dot{z}_\nu)-L\dot{z}_\nu\right]
\delta z^\nu\right|_{z_1}^{z_2}\\
\qquad+\int\limits_{z_1}^{z_2}d\tau\,\left[\frac{\partial
L}{\partial z^\mu}-\frac{d}{d\tau}\left\{ \frac{\partial
L}{\partial \dot{z}^\mu}(\delta^\mu{}_\nu+\dot{z}^\mu\dot{z}_\nu)
-L\dot{z}_\nu\right\}\right]\delta z^\nu.
\end{array}
\end{equation}
Taking that $\delta z^\nu_{1,2}\equiv\left.\delta
z^\nu\right|_{z_{1,2}}=0$, we get
\begin{equation}
\delta W\equiv\int\limits_{z_1}^{z_2}\frac{\delta W}{\delta
z^{\nu}}\delta z^{\nu}d\tau,
\end{equation}
where
\begin{equation}
\frac{\delta W}{\delta z^{\nu}}=\frac{\partial L}{\partial z^\nu}-
\frac{d}{d\tau}\left\{ \frac{\partial L}{\partial
\dot{z}^\mu}(\delta^\mu{}_\nu+\dot{z}^\mu\dot{z}_\nu)
-L\dot{z}_\nu\right\},
\end{equation}
and, by virtue of the stationary action principle, the following
equation of motion result from here
\begin{equation}
\frac{\partial L}{\partial z^\nu}- \frac{d}{d\tau}\left\{
\frac{\partial L}{\partial
\dot{z}^\mu}(\delta^\mu{}_\nu+\dot{z}^\mu\dot{z}_\nu)
-L\dot{z}_\nu\right\}=0.
\end{equation}
We come just to the Kalman equations [1]. Obviously, they differ
from the Euler - Lagrange equations by the terms resulted from the
manifest variation of the propertime element.
\subsection{Equations of
motion of the particle in the reparametrization-invariant
theories}

Parametrization of the world line of the particle by propertime
is too specific. It appears frequently convenient to refuse from
the above choice and to numerate the sequence of trajectory
points in somehow different manner. In this case, however, the
parameter $\theta$ used for such numeration will not have the
clear physical sense as $\tau$ has, and probably, will not be
observable at all. In addition, it may happen that it is
convenient to analyze some of the properties of the system at one
parametrization choice, while other properties of this system are
easier made apparent at the different parameter choice. Therefore
it seems expedient to construct initially the theory of point
particle interaction with the field in such a way to make it
independent on arbitrariness in the particle parametrization, i.e.
it should be reparametrizationally invariant.

Let the world line of the particle be parametrized by an arbitrary
Lorentz-invariant parameter $\theta$ with length dimension, i.e.
$z^\mu=z^\mu(\theta)$. Let us denote
\begin{equation}
z'^\mu\equiv\frac{dz^\mu(\theta)}{d\theta},\quad
v^\mu(\theta)\equiv\frac{z'^\mu(\theta)}{\sqrt{-z'^\alpha(\theta)
z'{}_\alpha(\theta)}}.
\end{equation}
If the Lagrange function of the particle depends on the
$z(\theta)$ and its derivatives over $\theta$ not higher than the
first order (and we will restrict ourselves to the consideration
of just such cases), then the action for the particle that
possesses reparametrization invariance {\it mast have} the
following form
\begin{equation}
W[z]=\int\limits\limz L(z(\theta),v(\theta))
\sqrt{-z'^\alpha(\theta) z'{}_\alpha(\theta)}d\theta.
\end{equation}
Action (1.18) is indeed reparametrization-invariant, since
$v^\mu$ and $\sqrt{-z'^\alpha(\theta)
z'{}_\alpha(\theta)}d\theta$ do not vary with reparametrization
transformations.

Let us find the equation of motion that follows from the action
functional (1.18). Let the parameter $\theta$ {\it be not varying
by definition}. Then
\begin{equation}
\delta W=\int\limits\limz\delta L \sqrt{-z'^\alpha
z'_\alpha}d\theta+ \int\limits\limz L \delta\sqrt{-z'^\alpha
z'_\alpha}d\theta,
\end{equation}
\begin{equation}
\delta L=\frac{\partial L}{\partial z^\mu}\delta z^\mu+
\frac{\partial L}{\partial z'^\mu}\delta z'^\mu= \frac{\partial
L}{\partial z^\mu}\delta z^\mu+ \frac{\partial L}{\partial
v^\mu}\frac{\partial v^\mu}{\partial z'^\nu}\delta z'^\nu.
\end{equation}
Taking into account that
\begin{equation} \frac{\partial v^\mu}{\partial
z'^\nu}= \frac{1}{\sqrt{-z'^\alpha
z'_\alpha}}(\delta^\mu{}_\nu+v^\mu v_\nu),
\end{equation}
and
\begin{equation}
\delta z'^\nu=\delta\left(\frac{dz^\nu}{d\theta}\right)=
\frac{d}{d\theta}\delta z^\nu,
\end{equation}
we find
\begin{equation}
\delta L=\frac{\partial L}{\partial z^\mu}\delta z^\mu+
\frac{\partial L}{\partial v^\mu}\frac{1}{\sqrt{-z'^\alpha
z'_\alpha}}(\delta^\mu{}_\nu+v^\mu v_\nu)\frac{d}{d\theta}\delta
z^\nu.
\end{equation}
Furthermore,
\begin{equation}
\delta\sqrt{-z'^\alpha z'_\alpha}=-v_\nu\frac{d}{d\theta}\delta
z^\nu.
\end{equation}
Substituting (1.23) and (1.24) into (1.19) and integrating by
parts, we have
\begin{equation}
\begin{array}{l}
\delta W=\int\limits\limz\left[\sqrt{-z'^\alpha
z'_\alpha}\frac{\partial L}{\partial z^\mu}\delta z^\mu+
\frac{\partial L}{\partial v^\mu}(\delta^\mu{}_\nu+v^\mu v_\nu)
\frac{d}{d\theta}\delta z^\nu- Lv_\mu \frac{d}{d\theta}\delta z^\nu\right]d\theta\\
\qquad=\left.\left[\frac{\partial L}{\partial v^\mu}
(\delta^\mu{}_\nu+v^\mu v_\nu)-Lv_\nu\right]
\delta z^\nu\right|\limz\\
\qquad+\int\limits\limz\left[\sqrt{-z'^\alpha
z'_\alpha}\frac{\partial L}{\partial z^\nu}-
\frac{d}{d\theta}\left\{ \frac{\partial L}{\partial
v^\mu}(\delta^\mu{}_\nu+v^\mu v_\nu)- Lv_\nu\right\}\right]\delta
z^\nu d\theta.
\end{array}
\end{equation}
Considering, as usually, that at the integration boundary
$\left.\delta z_\nu\right|_{\theta_{1,2}}=0$, and that the action
is extremal at the real trajectories, we find from (1.25) the
following equations of motion
\begin{equation}
\frac{\delta W}{\delta z^\nu}=\sqrt{-z'^\alpha
z'_\alpha}\frac{\partial L}{\partial
z^\nu}-\frac{d}{d\theta}\left\{\frac{\partial L}{\partial
v^\mu}\left(\delta^\mu{}_\nu+v^\mu v_\nu\right)-Lv_\nu\right\}=0,
\end{equation}
or, after dividing by $\sqrt{-z'^\alpha z'_\alpha}$,
\begin{equation}
\frac{\partial L}{\partial z^\nu}-\frac{1}{\sqrt{-z'^\alpha
z'_\alpha}}\frac{d}{d\theta}\left\{\frac{\partial L}{\partial
v^\mu}\left(\delta^\mu{}_\nu+v^\mu v_\nu\right)-Lv_\nu\right\}=0.
\end{equation}
It is easy to see that equations (1.27) are the RI-equations.

Let us choose the propertime of the particle as the parameter
\begin{equation}
\tau=\int\limits_{\tau_0}^{\tau}\sqrt{-z'^\alpha
z'_\alpha}d\theta,
\end{equation}
where $\tau_0$ is arbitrary. Then $z'^\alpha$ is transformed into
$\dot{z}^\alpha$, $\dot{z}^\alpha\dot{z}_\alpha=-1$, and $v^\mu$
into $\dot{z}^\mu$. Hence, in the propertime gauge, equations
(1.27) take a form
\begin{equation}
\frac{\partial L}{\partial z^\nu}- \frac{d}{d\tau}\left\{
\frac{\partial L}{\partial
\dot{z}^\mu}(\delta^\mu{}_\nu+\dot{z}^\mu\dot{z}_\nu)
-L\dot{z}_\nu\right\}=0,
\end{equation}
that exactly coincides with the Kalman equations (1.16).

Thus, the Kalman equations derived with the help of manifest
variation of the particle propertime element can be reproduced
also as the Euler-Lagrange equations in the RI-theory in the
propertime gauge.

Let us introduce the following notations to reach manifest RI for
all relations:
\begin{equation}
\dt\equiv\sqrt{-z'^\alpha z'_\alpha}d\theta
\end{equation}
is the RI-"volume element"\,(RI-integration measure);
\begin{equation}
D\equiv\frac{1}{\sqrt{-z'^\alpha z'_\alpha}}\frac{d}{d\theta}
\end{equation}
is the RI-derivative. Then
\begin{equation}
v^\mu=Dz^\mu
\end{equation}
and (1.18), (1.25) and (1.26) will be written as
\begin{equation}
W[z]=\intz L(z(\theta),Dz(\theta)),
\end{equation}
\begin{equation}
\begin{array}{l}
\delta W=\left.\left[\frac{\partial L}{\partial Dz^\mu}
(\delta^\mu{}_\nu+Dz^\mu Dz_\nu)-LDz_\nu\right]
\delta z^\nu\right|\limz\\
\quad+\intz\left[\frac{\partial L}{\partial z^\nu}- D\left\{
\frac{\partial L}{\partial Dz^\mu}(\delta^\mu{}_\nu+Dz^\mu
Dz_\nu)- LDz_\nu\right\}\right]\delta z^\nu,
\end{array}
\end{equation}
\begin{equation}
\frac{\Delta W}{\Delta z^\nu}=\frac{\partial L}{\partial z^\nu}-
D\left\{ \frac{\partial L}{\partial
Dz^\mu}(\delta^\mu{}_\nu+Dz^\mu Dz_\nu)- LDz_\nu\right\}=0,
\end{equation}
where
\begin{equation}
\frac{\Delta W}{\Delta z^\nu}\equiv\frac{1}{\sqrt{-z'^\alpha
z'_\alpha}}\frac{\delta W}{\delta z^\nu}
\end{equation}
is the RI-variation derivative. In the propertime gauge,
$\dt\rightarrow d\tau$, $D\rightarrow d/d\tau$,
$Dz^\mu\rightarrow\dot{z}^\mu$.

Let us show also the RI-analog of the formula for integration by
parts:
\begin{equation}
\intz\left(Df(\theta)\right)g(\theta)=
\left.\left[f(\theta)g(\theta)\right]\right|\limz- \intz
f(\theta)\left(Dg(\theta)\right).
\end{equation}
In particular,
\begin{equation}
\intz\left(Df(\theta)\right)= \left.f(\theta)\right|\limz.
\end{equation}
\subsection{N\H{o}ether identity. Effective mass and force}

The RI can be considered as invariance with respect to the local
Abel group of transformations
\begin{equation}
\left\{
\begin{array}{l}
\theta\rightarrow\tilde\theta=\theta+\delta\varepsilon(\theta)\\
z^\mu(\theta)\rightarrow\tilde{z}^\mu(\tilde{\theta})=z^\mu(\theta),
\end{array}
\right.
\end{equation}
where $\delta\varepsilon(\theta)$ is an arbitrary infinitesimal
function –- the transformation parameter. We find from the second
formula of (1.39)
\begin{equation}
\bar{\delta}z^\mu(\theta)\equiv\tilde{z}^\mu(\tilde{\theta})-z^\mu(\theta)=0
\end{equation}
and, since
\begin{equation}
\bar{\delta}z^\mu(\theta)=\delta z^\mu(\theta)+
z'^\mu(\theta)\delta\varepsilon(\theta)=\delta z^\mu(\theta)+
Dz^\mu(\theta)\Delta\varepsilon(\theta),
\end{equation}
where
\begin{equation} \delta
z^\mu(\theta)\equiv\tilde{z}^\mu(\theta)-z^\mu(\theta),
\end{equation}
\begin{equation}
\Delta\varepsilon(\theta)\equiv\sqrt{-z'^\alpha z'_\alpha}
\delta\varepsilon(\theta)
\end{equation}
is the RI-variation of parameter, then at the transformations
(1.39)
\begin{equation}
\delta z^\mu(\theta)=-Dz^\mu(\theta)\Delta\varepsilon(\theta).
\end{equation}
On the strength of the second N\H{o}ether theorem, the following
identity takes place
\begin{equation}
\frac{\Delta W}{\Delta z^\mu}Dz^\mu=0,
\end{equation}
that means that the Eulerian $\Delta W/\Delta z^\nu$ has the
following structure
\begin{equation}
\frac{\Delta W}{\Delta z^\nu}=f_\mu(\delta^\mu{}_\nu+Dz^\mu
Dz_\nu),
\end{equation}
where $f_\mu$ is a some RI-4-vector. The quantity
\begin{equation} h^\mu_\nu\equiv\delta^\mu{}_\nu+Dz^\mu
Dz_\nu
\end{equation}
is the projector onto the proper 3-space of the particle, since
the above quantity coincides with it in the propertime gauge and
is the RI-quantity.

Let us prove formula (1.46) using an manifest form of the
variation derivation $\Delta W/\Delta z^\nu$ (1.35). Using (1.31)
and (1.32), it is easy to derive the auxiliary formulae
\begin{equation}
D^2 z^\mu\equiv DDz^\mu=(\delta^\mu{}_\nu+Dz^\mu
Dz_\nu)\frac{z''^\nu}{(\sqrt{-z'^\alpha z'_\alpha})^2},
\end{equation}
\begin{equation}
Df(z(\theta),Dz(\theta))=\frac{\partial f(z,Dz)}{\partial
z^\mu}Dz^\mu+\frac{\partial f(z,Dz)}{\partial Dz^\mu}D^2 z^\mu,
\end{equation}
where $f(z(\theta),Dz(\theta))$ is an  arbitrary differentiable
function of the above arguments. Formula (1.49) is the RI-analog
of the known analysis formula for a differentiation of composite
function.

Using (1.35) and (1.49), we find
\begin{equation}
\begin{array}{l} \frac{\Delta W}{\Delta z^\nu}=\frac{\partial L}{\partial
z^\nu}- D\left\{ \frac{\partial L}{\partial
Dz^\mu}(\delta^\mu{}_\nu+Dz^\mu Dz_\nu)-
LDz_\nu\right\}=\frac{\partial L}{\partial z^\nu}- D\left(
\frac{\partial L}{\partial Dz^\mu}\right)(\delta^\mu{}_\nu+Dz^\mu Dz_\nu)\\
\qquad-\frac{\partial L}{\partial Dz^\mu}(D^2 z^\mu Dz_\nu+Dz^\mu
D^2z_\nu)+\left(\frac{\partial L}{\partial z^\mu}Dz^\mu+
\frac{\partial L}{\partial Dz^\mu}D^2
z^\mu\right)Dz_\nu+LD^2z_\nu\\
\qquad=\left(\frac{\partial L}{\partial z^\mu}-D\frac{\partial
L}{\partial Dz^\mu}\right)(\delta^\mu{}_\nu+Dz^\mu Dz_\nu)+
\left(L-\frac{\partial L}{\partial Dz^\mu}Dz^\mu\right)D^2 z_\mu.
\end{array}
\end{equation}
Taking into account the formula
\begin{equation}
D^2 z^\mu Dz_\mu=0,
\end{equation}
that follows from the identity
\begin{equation}
Dz^\mu Dz_\mu=-1,
\end{equation}
we obtain finally
\begin{equation}
\begin{array}{l}
\frac{\Delta W}{\Delta z^\nu}=\left[ \left(\frac{\partial
L}{\partial z^\mu}-D\frac{\partial L}{\partial Dz^\mu}\right)-
\left(\frac{\partial L}{\partial Dz^\alpha}Dz^\alpha-L\right)D^2
z_\mu\right](\delta^\mu{}_\nu+Dz^\mu Dz_\nu),
\end{array}
\end{equation}
that coincides with (1.46) at
\begin{equation}
f_\mu=\left(\frac{\partial L}{\partial z^\mu}-D\frac{\partial
L}{\partial Dz^\mu}\right)- \left(\frac{\partial L}{\partial
Dz^\alpha}Dz^\alpha-L\right)D^2 z_\mu.
\end{equation}
Now the equations of motion $\Delta W/\Delta z^\nu=0$ can be
written as
\begin{equation}
\left(\frac{\partial L}{\partial Dz^\alpha}Dz^\alpha-L\right)D^2
z_\nu=\left(\frac{\partial L}{\partial z^\mu}-D\frac{\partial
L}{\partial Dz^\mu}\right)(\delta^\mu{}_\nu+Dz^\mu Dz_\nu).
\end{equation}
It is natural to call
\begin{equation}
M\equiv\frac{\partial L}{\partial Dz^\alpha}Dz^\alpha-L
\end{equation}
the effective (dynamical) mass of the particle, and
\begin{equation}
F_\nu\equiv\left(\frac{\partial L}{\partial z^\mu}-D\frac{\partial
L}{\partial Dz^\mu}\right)(\delta^\mu{}_\nu+Dz^\mu Dz_\nu)
\end{equation}
the effective 4-force that the field acts on the particle with.
In the above notations the equations of motion will take a simple
form
\begin{equation}
MD^2z_\nu=F_\nu.
\end{equation}
\section{Total system "particle+field"}
\setcounter{equation}{0}
\subsection{Total action functional variation. Equations of motion of
particle and field}

Consider now the total system that consists of interacting point
relativistic particle and certain set of fields and is described
by the action functional of the following form
\begin{equation}
\begin{array}{l}
W[z,\phi;\theta_{1,2},\Sigma_{1,2}]=W^P[z;\theta_{1,2}]+W^I[z,\phi;\theta_{1,2},\Sigma_{1,2}]+W^F[\phi;\Sigma_{1,2}]\\
\qquad=\intz L^P(z(\theta),Dz(\theta))+ \intz\intx
L^I(z(\theta),Dz(\theta);\phi(x),\partial\phi(x))\delta(x-z(\theta))\\
\qquad+\intx \Lag ^F(\phi(x),\partial\phi(x)).
\end{array}
\end{equation}
If one requires the Lagrange functions $L^{P,I}$ to be
form-invariant scalars with respect to the transformations of the
Poincair\'{e} group, then the condition $\partial L^{P,I}/\partial
z^\mu=0$ must hold true. For the Lagrange function of the particle
this means $L^P=L^P(Dz)$. However, on the strength of identity
$Dz^\mu Dz_\mu=-1$, it is impossible to construct nontrivial
scalar only of $Dz^\mu$. Therefore the only possibility remains:
$L^P=C=const$. Usually $C=-m$ is taken, where $m$ is the
(kinematical) mass of the particle. Thus, the action functional
for the total relativistic system "particle+field"\,must have a
form
\begin{equation}
\begin{array}{l}
W=W^P+W^I+W^F\\
\qquad=-\intz m+ \intz\intx
L^I(Dz(\theta);\phi(x),\partial\phi(x))\delta(x-z(\theta))+\intx
\Lag ^F(\phi(x),\partial\phi(x)).
\end{array}
\end{equation}
To derive the equations of motion and to construct dynamical
variables we need an expression for the total variation of the
action functional (2.2)
\begin{equation}
\bar{\delta}W[z,\phi;\theta_{1,2},\Sigma_{1,2}]\equiv W[z+\delta z
,\phi+\delta\phi;\theta_{1,2}+\delta\theta_{1,2},\Sigma_{1,2}+\delta\Sigma_{1,2}]-
W[z,\phi;\theta_{1,2},\Sigma_{1,2}].
\end{equation}
Consider separately the total variations for each of three terms
in (2.2)
\begin{equation}
\bar{\delta}W^P[z;\theta_{1,2}]=\left.[-m\triangle\theta]\right|_{\theta_1}^{\theta_2}+\delta_zW^P,
\end{equation}
where $\triangle\theta\equiv\sqrt{-z'^\alpha(\theta)
z'{}_\alpha(\theta)}\delta\theta$.
\begin{equation}
\begin{array}{l}
\delta_z W^P[z;\theta_{1,2}]\equiv W^P[z+\delta
z;\theta_{1,2}]-W^P[z;\theta_{1,2}]\\
\qquad=-\int\limits_{\theta_1}^{\theta_2}\delta_z(\dt)\,m= \intz m
Dz_\nu D(\delta z^\nu)=\left.[m Dz_\nu\delta
z^\nu]\right|_{\theta_1}^{\theta_2}- \intz m D^2z_\nu \delta
z^\nu.
\end{array}
\end{equation}
When deriving (2.5) we have used formula $\delta_z(\dt)=-\dt
Dz_\nu D(\delta z^\nu)$ and performed integration by parts.
Substituting (2.5) into (2.4), we find
\begin{equation}
\bar{\delta}W^P=\left.[-m\triangle\theta+m Dz_\nu\delta
z^\nu]\right|_{\theta_1}^{\theta_2}- \intz m D^2z_\nu \delta
z^\nu.
\end{equation}

For pure field term, as usually,
\begin{equation}
\bar{\delta}W^F[\phi;\Sigma_{1,2}]\equiv
W^F[\phi+\delta\phi;\Sigma_{1,2}+\delta\Sigma_{1,2}]-W^F[\phi;\Sigma_{1,2}]=
\left.\left[\ints\Lag ^F\delta x^\mu\right]\right|\limx
+\delta_\phi W^F,
\end{equation}
where
\begin{equation}
\begin{array}{l}
\delta_\phi W^F[\phi;\Sigma_{1,2}]\equiv
W^F[\phi+\delta\phi;\Sigma_{1,2}]-W^F[\phi;\Sigma_{1,2}]\\
\qquad=\intx \delta_\phi\Lag ^F= \int\limits\limx dx
\left[\frac{\partial\Lag ^F}{\partial\phi}\delta\phi+
\frac{\partial\Lag ^F}{\partial\phi,_\mu}\delta(\phi,_\mu)\right]=
\left.\left[\int\limits_\Sigma d\sigma_\mu\,\frac{\partial\Lag
^F}{\partial\phi,_\mu}\delta \phi\right]\right|\limx+ \intx
\left[\frac{\partial\Lag ^F}{\partial\phi}-\partial_\mu
\frac{\partial\Lag ^F}{\partial\phi,_\mu}\right]\delta\phi.
\end{array}
\end{equation}
Substituting (2.8) into (2.7), we get
\begin{equation}
\bar{\delta}W^F=\left.\left[\int\limits_\Sigma d\sigma_\mu\,\Lag
^F\delta x^\mu+\int\limits_\Sigma d\sigma_\mu\,\frac{\partial\Lag
^F}{\partial\phi,_\mu}\delta \phi\right]\right|\limx+ \intx
\left[\frac{\partial\Lag ^F}{\partial\phi}-\partial_\mu
\frac{\partial\Lag ^F}{\partial\phi,_\mu}\right]\delta\phi.
\end{equation}
Furthermore,
\begin{equation}
\begin{array}{l}
\bar{\delta}W^I[z,\phi;\theta_{1,2},\Sigma_{1,2}]\equiv
W^I[z+\delta z,\phi+\delta\phi;\theta_{1,2}+\delta\theta_{1,2},
\Sigma_{1,2}+\delta\Sigma_{1,2}]-W^I[z,\phi;\theta_{1,2},\Sigma_{1,2}]\\
\qquad=\left.\left[\left(\intx L^I\xz
\right)\triangle\theta\right]
\right|_{\theta_1}^{\theta_2}+\left.\left[\int\limits_\Sigma
d\sigma_\mu\left(\intz L^I\xz \right)\delta
x^\mu\right]\right|\limx+\delta_z W^I+\delta_\phi W^I,
\end{array}
\end{equation}
\begin{equation}
\begin{array}{l}
\delta_z W^I[z,\phi;\theta_{1,2},\Sigma_{1,2}]\equiv W^I[z+\delta
z,\phi;\theta_{1,2},\Sigma_{1,2}]-W^I[z,\phi;\theta_{1,2},\Sigma_{1,2}]\\
\qquad=\int\limits_{\theta_1}^{\theta_2}\delta_z(\dt)\,\intx
L^I\xz +
\intz\intx \delta_z(L^I\xz )\\
\qquad=-\intz\intx L^I Dz_\nu \xz D(\delta z^\nu)+ \intz\intx
\delta_z L^I\xz + \intz\intx L^I\delta_z\xz.
\end{array}
\end{equation}
Using formulae
\begin{equation}
\begin{array}{l}
\delta_z(\dt)=-\dt Dz_\nu D(\delta z^\nu),\quad\delta_z
f(z,Dz)=\frac{\partial f}{\partial
z^\nu}\delta z^\nu+ \frac{\partial f}{\partial Dz^\nu}\delta_z (Dz^\nu),\\
\delta_z (Dz^\mu)=(\delta^\mu{}_\nu+Dz^\mu Dz_\nu)D(\delta z^\nu),
\end{array}
\end{equation}
and integrating by parts, we find for the first two terms
\begin{equation}
\begin{array}{l}
\left.\left[\intx \left\{\imp{\mu}{\nu}\right\}\xz \delta z^\nu\right]\right|_{\theta_1}^{\theta_2}\\
\qquad-\intz D\left[\intx \left\{\imp{\mu}{\nu}\right\}\xz
\right]\delta z^\nu.
\end{array}
\end{equation}
For the third term
\begin{equation}
\begin{array}{l}
\intz\intx L^I\delta_z\xz =\intz\intx L^I\frac{\partial\xz
}{\partial z^\nu}\delta z^\nu=-\intz\intx
L^I\frac{\partial\xz }{\partial x^\nu}\delta z^\nu\\
\qquad=\left.\left[-\int\limits_\Sigma d\sigma_\nu\left(\intz
L^I\xz \delta z^\nu\right)\right]\right|\limx+\intz \intx
\frac{dL^I}{dx^\nu}\xz \delta z^\nu,
\end{array}
\end{equation}
where
\begin{equation}
\frac{dL^I}{dx^\nu}\equiv\frac{\partial
L^I}{\partial\phi(x)}\phi,_\nu(x)+\frac{\partial
L^I}{\partial\phi,_\mu(x)}\phi,_{\mu\nu}(x).
\end{equation}
Compiling together, we obtain
\begin{equation}
\begin{array}{l}
\delta_z W^I= \left.\left[\intx \left\{\frac{\partial
L^I}{\partial Dz^\mu}(\delta^\mu{}_\nu+Dz^\mu Dz_\nu)-L^I
Dz_\nu\right\}\xz \delta z^\nu\right]\right|_{\theta_1}^{\theta_2}\\
\qquad-\left.\left[\int\limits_\Sigma d\sigma_\nu\left(\intz
L^I\xz \delta
z^\nu\right)\right]\right|\limx\\
\qquad+\intz\left[-D\left(\intx \left\{\imp{\mu}{\nu}\right\}\xz
\right)+\intx \frac{dL^I}{dx^\nu}\xz \right]\delta z^\nu,
\end{array}
\end{equation}
Furthermore
\begin{equation}
\begin{array}{l}
\delta_\phi W^I[z,\phi;\theta_{1,2},\Sigma_{1,2}]\equiv
W^I[z,\phi+\delta\phi;\theta_{1,2},\Sigma_{1,2}]-W^I[z,\phi;\theta_{1,2},\Sigma_{1,2}]\\
\qquad=\intz\intx  \delta_\phi L^I\xz = \intz\intx
\left[\frac{\partial L^I}{\partial\phi(x)}\delta\phi(x)+
\frac{\partial L^I}{\partial\phi,_\mu(x)}\delta_\phi(\phi,_\mu(x))
\right]\xz \\
\qquad=\left.\left[\ints\left(\intz\frac{\partial
L^I}{\partial\phi,_\mu(x)}\xz \right)\delta\phi(x)\right]\right|\limx\\
\qquad+\intz\intx\left[\frac{\partial L^I}{\partial\phi(x)}\xz
-\frac{\partial}{\partial x^\mu}\left(\frac{\partial
L^I}{\partial\phi,_\mu(x)}\xz \right) \right]\delta\phi(x).
\end{array}
\end{equation}
Substituting (2.16), (2.17) into (2.10)
\begin{equation}
\begin{array}{l}
\bar\delta W^I=\left.\left[\intx L^I\xz
\triangle\theta+\intx\left\{\imp{\mu}{\nu}\right\}\xz \delta
z^\nu\right]\right|\limz\\
\qquad+\left.\left[\ints\intz L^I\xz(\delta x^\mu-\delta
z^\mu)+\ints\left(\intz\frac{\partial
L^I}{\partial\phi,_\mu}\xz\right)\delta\phi\right]\right|\limx\\
\qquad+\intz\left[-D\left(\intx\left\{\imp{\mu}{\nu}\right\}\xz\right)+
\intx\frac{dL^I}{dx^\nu}\xz\right]\delta z^\nu\\
\qquad\,\,\,+\intx\left[\intz\frac{\partial
L^I}{\partial\phi(x)}\xz- \frac{\partial}{\partial
x^\mu}\left(\intz\frac{\partial
L^I}{\partial\phi,_\mu(x)}\xz\right)\right]\delta\phi.
\end{array}
\end{equation}
The first term in the second square parenthesis equals to zero,
since at any variation of the particle coordinate the equality
$\delta x^\nu=\delta z^\nu$ holds true. Validity of the last
statement results from the following simple consideration. Any
function (or functional) of the particle coordinate $f(z)$ can be
presented in a form
\begin{equation}
f(z)=\intx f(x)\xz\equiv F(z;\Sigma_{1,2}).
\end{equation}
Let us take variation of the both parts of (2.19). On the
left-hand side we have
\begin{equation}
\delta_z f(z)=\frac{\partial f(z)}{\partial z^\mu}\delta z^\mu,
\end{equation}
whereas on the right-hand side
\begin{equation}
\bar\delta F(z;\Sigma_{1,2})\equiv F(z+\delta
z;\Sigma_{1,2}+\delta\Sigma_{1,2})-F(z;\Sigma_{1,2})=
\left.\left[\ints f(x)\xz\delta x^\mu\right]
\right|\limx+\delta_zF,
\end{equation}
where
\begin{equation}
\begin{array}{l}
\delta_z F(z;\Sigma_{1,2})\equiv F(z+\delta
z;\Sigma_{1,2})-F(z;\Sigma_{1,2})\\
\qquad=\intx f(x)\frac{\partial\xz}{\partial z^\mu}\delta z^\mu=
-\intx f(x)\frac{\partial\xz}{\partial x^\mu}\delta z^\mu\\
\qquad=\intx\left[-\frac{\partial}{\partial
x^\mu}\left(f(x)\xz\right)\delta z^\mu+\frac{\partial
f(x)}{\partial x^\mu}\xz\delta z^\mu\right]\\
\qquad=\left.\left[\ints f(x)\xz\delta
z^\mu\right]\right|\limx+\frac{\partial f(z)}{\partial
z^\mu}\delta z^\mu.
\end{array}
\end{equation}
Substituting (2.22) into (2.21) and comparing the result with
(2.20), we find
\begin{equation}
\left.\left[\ints f(x)\xz(\delta z^\mu-\delta
x^\mu)\right]\right|\limx=0,
\end{equation}
and, on the strength of arbitrariness of the function $f(z)$ and
hypersurfaces $\Sigma_{1,2}$,
\begin{equation} \delta x^\mu=\delta z^\mu.
\end{equation}
Let us denote
\begin{equation}
\pi_\nu\equiv\intx\left\{\imp{\mu}{\nu}\right\}
\end{equation}
\begin{equation}
\Lag ^I\equiv\intz L^I\xz.
\end{equation}
Then, compiling (2.6), (2.9) and (2.18), we have
\begin{equation}
\begin{array}{l}
\bar{\delta}W=\left.\left[\left(-m+\intx
L^I\xz\right)\triangle\theta+(mDz_\nu+\pi_\nu)\delta z^\nu
\right]\right|\limz\\
\qquad+\left.\left[\ints\Lag ^F\delta x^\mu+
\ints\frac{\partial(\Lag ^I+\Lag ^F)}{\partial\phi,_\mu}\delta\phi
\right]\right|\limx\\
\qquad+\intz\left[-D(mDz_\nu+\pi_\nu)+\intx\frac{DL^I}{dx^\nu}\xz\right]\delta
z^\nu\\
\qquad+\intx\left[\frac{\partial(\Lag^F+\Lag^I)}{\partial\phi}-
\frac{\partial}{\partial
x^\mu}\left(\frac{\partial(\Lag^F+\Lag^I)}{\partial\phi,_\mu}\right)\right]\delta\phi.
\end{array}
\end{equation}
As usually, the boundary terms in (2.27) are expressed via the
total variations $\bar{\delta}z^\nu$, $\bar{\delta}\phi$ of the
coordinate of the particle and field function, respectively. If
as a result of variation
\begin{equation}
\begin{array}{l}
\theta\rightarrow\tilde{\theta}\equiv\theta+\delta\theta,\quad
z^\nu(\theta)\rightarrow\tilde{z}^\nu(\theta)\equiv
z^\nu(\theta)+\delta z^\nu(\theta),\\
x\rightarrow\tilde x\equiv x+\delta x,\quad
\phi(x)\rightarrow\tilde\phi(x)\equiv\phi(x)+\delta\phi(x),
\end{array}
\end{equation}
then
\begin{equation}
\begin{array}{l}
\bar{\delta}z^\nu(\theta)\equiv\tilde{z}^\nu(\tilde{\theta})-z^\nu(\theta)=
\delta z^\nu(\theta)+Dz^\nu(\theta)\triangle\theta,\\
\bar{\delta}\phi(x)\equiv\tilde{\phi}(\tilde{x})-\phi(x)= \delta
\phi(x)+\phi,_\nu(x)\delta x^\nu(x),
\end{array}
\end{equation}
hence
\begin{equation}
\delta z^\nu(\theta)=\bar{\delta}z^\nu(\theta)-
Dz^\nu(\theta)\triangle\theta,\quad \delta
\phi(x)=\bar{\delta}\phi(x)-\phi,_\nu(x)\delta x^\nu(x).
\end{equation}
Substituting (2.30) into (2.27) and taking into account (2.24),
we find finally
\begin{equation}
\begin{array}{l}
\bar{\delta}W=\left.\left[(mDz_\nu+\pi_\nu)\bar{\delta}z^\nu\right]\right|\limz\\
\qquad+\left.\left[-\ints\left\{\frac{\partial(\Lag^F+\Lag^I)}{\partial\phi,_\mu}\phi,_\nu-
\Lag^F\delta^\mu{}_\nu\right\}\delta
x^\nu+\ints\frac{\partial(\Lag^F+\Lag^I)}{\partial\phi,_\mu}\bar{\delta}\phi
\right]\right|\limx\\
\qquad+\intz\left[-D(mDz_\nu+\pi_\nu)+\intx\frac{dL^I}{dx^\nu}\xz\right]\delta z^\nu\\
\qquad+\intx\left[\frac{\partial(\Lag^F+\Lag^I)}{\partial\phi}-
\frac{\partial}{\partial
x^\mu}\left(\frac{\partial(\Lag^F+\Lag^I)}{\partial\phi,_\mu}\right)\right]\delta\phi.
\end{array}
\end{equation}
The following equations of motion for the particle and fields
result from (2.31)
\begin{equation}
\frac{\triangle W}{\triangle z^\nu}=
-D(mDz_\nu+\pi_\nu)+\intx\frac{dL^I}{dx^\nu}\xz=0,
\end{equation}
\begin{equation}
\frac{\delta W}{\delta\phi(x)}=
\frac{\partial(\Lag^F+\Lag^I)}{\partial\phi}-
\frac{\partial}{\partial
x^\mu}\left(\frac{\partial(\Lag^F+\Lag^I)}{\partial\phi,_\mu}\right)=0.
\end{equation}
At the equations of motion
\begin{equation}
\begin{array}{l}
\bar{\delta}W=\left.\left[(mDz_\nu+\pi_\nu)\bar{\delta}z^\nu\right]\right|\limz\\
\qquad+\left.\left[-\ints\left\{\frac{\partial(\Lag^F+\Lag^I)}{\partial\phi,_\mu}\phi,_\nu-
\Lag^F\delta^\mu{}_\nu\right\}\delta
x^\nu+\ints\frac{\partial(\Lag^F+\Lag^I)}{\partial\phi,_\mu}\bar{\delta}\phi
\right]\right|\limx.
\end{array}
\end{equation}
Formula (2.34) makes a basis for constructing the dynamical
variables of the system "particle+field". Note that variation of
the time parameter $\triangle\theta$ is not involved in (2.34).
\subsection{4-Momentum vector. Energy-momentum tensor}

At the space-time translations
\begin{equation}
\left\{\begin{array}{l} \delta\theta=0\\
\delta z^\mu=\delta x^\mu=\delta\varepsilon^\mu,\,\delta\varepsilon^\mu=\mathrm{const}\\
\bar{\delta}\phi=0
\end{array}\right.
\end{equation}
and formula (2.34) provides an expression for the 4-momentum
vector of the system:
\begin{equation}
\bar{\delta}W=\left.[P_\nu[\Sigma]]\right|\limx\delta\varepsilon^\nu.
\end{equation}
Hence,
\begin{equation}
P_\nu[\Sigma]=(mDz_\nu+\pi_\nu)-
\ints\left\{\frac{\partial(\Lag^F+\Lag^I)}{\partial\phi,_\mu}\phi,_\nu-
\Lag^F\delta^\mu{}_\nu\right\},\qquad(z(\theta)\in\Sigma).
\end{equation}
The 4-momentum vector can also be expressed via the
energy-momentum tensor (EMT) by the following formula
\begin{equation}
P_\nu[\Sigma]=\ints T^\mu{}_\nu.
\end{equation}
Then
\begin{equation}
\begin{array}{l}
T^\mu{}_\nu=\intzi'Dz^\mu(mDz_\nu+\pi_\nu)\xz-
\left\{\frac{\partial(\Lag^F+\Lag^I)}{\partial\phi,_\mu}\phi,_\nu-
\Lag^F\delta^\mu{}_\nu\right\}\\
\qquad=\left\{\intzi'mDz^\mu Dz_\nu\xz\right\}+
\left\{\intzi'Dz^\mu\pi_\nu\xz+\frac{\partial\Lag^I}{\partial\phi,_\mu}\phi_\nu\right\}\\
\qquad-\left\{\frac{\partial\Lag^F}{\partial\phi,_\mu}\phi,_\nu-
\Lag^F\delta^\mu{}_\nu\right\}\equiv
T^P{}^\mu{}_\nu+T^I{}^\mu{}_\nu+T^F{}^\mu{}_\nu.
\end{array}
\end{equation}
Let us show that EMT (2.39) is conservative.
\begin{equation}
\begin{array}{l}
\partial_\mu T^\mu{}_\nu=\intzi'
Dz^\mu(mDz_\nu+\pi_\nu)\frac{\partial\xz}{\partial x^\mu}\\
\qquad-\left\{\partial_\mu\left(\frac{\partial(\Lag^F+\Lag^I)}{\partial\phi,_\mu}\right)\phi,_\nu+
\frac{\partial(\Lag^F+\Lag^I)}{\partial\phi,_\mu}\phi,_{\mu\nu}-\partial_\nu\Lag^F\right\}=
-\intzi'Dz^\mu(mDz_\nu+\pi_\nu)\frac{\partial\xz}{\partial
z^\mu}\\
\qquad-\left\{\partial_\mu\left(\frac{\partial(\Lag^F+\Lag^I)}{\partial\phi,_\mu}\right)\phi,_\nu+
\frac{\partial(\Lag^F+\Lag^I)}{\partial\phi,_\mu}\phi,_{\mu\nu}-
\frac{\partial\Lag^F}{\partial\phi}\phi,_\nu-
\frac{\partial\Lag^F}{\partial\phi,_\mu}\phi,_{\mu\nu}\right\}\\
\qquad=-\intzi'(mDz_\nu+\pi_\nu)D\xz-
\left[\frac{\partial\Lag^I}{\partial\phi}\phi,_\nu-
\frac{\partial\Lag^I}{\partial\phi,_\mu}\phi,_{\mu\nu}\right]\\
\qquad-\left[\partial_\mu\left(\frac{\partial(\Lag^F+\Lag^I)}{\partial\phi,_\mu}\right)-
\frac{\partial(\Lag^F+\Lag^I)}{\partial\phi}\right]\phi,_\nu=
\left.\left[-(mDz_\nu+\pi_\nu)\xz\right]\right|\limi\\
\qquad+\intzi'\left[D(mDz_\nu+\pi_\nu)-\frac{dL^I}{dx^\nu}\right]\xz+
\left[\frac{\partial(\Lag^F+\Lag^I)}{\partial\phi}-
\partial_\mu\left(\frac{\partial(\Lag^F+\Lag^I)}{\partial\phi,_\mu}\right)\right]\phi,_\nu.
\end{array}
\end{equation}
The first term for finite $x^0$ equals to zero, and the second and
third terms turn to zero at the equations of motion (2.32),
(2.33). Hence,
\begin{equation}
\partial_\mu T^\mu{}_\nu=0.
\end{equation}
Note that EMT (2.39) is not symmetrical: $T_{\mu\nu}\neq
T_{\nu\mu}$.
\subsection{4-Angular momentum tensor. 4-Angular momentum density tensor}

At the transformations from the Lorentz group (boosts and
rotations)
\begin{equation}
\left\{\begin{array}{l} \delta\theta=0\\
\delta z^\mu=\frac{1}{2}\delta\varepsilon^{\rho\sigma}
(z_\rho\delta^\mu{}_\sigma-z_\sigma\delta^\mu{}_\rho)\\
\delta x^\mu=\frac{1}{2}\delta\varepsilon^{\rho\sigma}
(x_\rho\delta^\mu{}_\sigma-x_\sigma\delta^\mu{}_\rho),
\quad\delta\varepsilon^{\rho\sigma}=\mathrm{const}\\
\bar{\delta}\phi=-\frac{1}{2}\frac{i}{\hbar}\delta\varepsilon^{\rho\sigma}
(S_{\rho\sigma})\phi,
\end{array}\right.
\end{equation}
In the last formula the truncated notations are used. The total
form is
$\bar{\delta}\phi^A=-\frac{1}{2}\frac{i}{\hbar}\delta\varepsilon^{\rho\sigma}
(S_{\rho\sigma})^A{}_B\phi^B$, where $A,B=\overline{1,N}$; $N$ is
the number of field components $\phi$. Then formula (2.34) gives
the expressions for the 4-angular momentum tensor of the system:
\begin{equation}
\bar{\delta}W=\frac{1}{2}\left.\left[J_{\rho\sigma}[\Sigma]\right]\right|\limx
\delta\varepsilon^{\rho\sigma}
\end{equation}
\begin{equation}
\begin{array}{l}
J_{\rho\sigma}[\Sigma]=\left\{z_\rho(mDz_\sigma+\pi_\sigma)-z_\sigma(mDz_\rho+\pi_\rho)\right\}\\
\quad-\ints\left\{x_\rho\left(\frac{\partial(\Lag^F+\Lag^I)}{\partial\phi,_\mu}\phi_\sigma-
\Lag^F\delta^\mu{}_\sigma\right)-x_\sigma\left(\frac{\partial(\Lag^F+\Lag^I)}{\partial\phi,_\mu}\phi_\rho-
\Lag^F\delta^\mu{}_\rho\right)\right\}-
\frac{i}{\hbar}\ints\frac{\partial(\Lag^F+\Lag^I)}{\partial\phi,_\mu}S_{\rho\sigma}\phi.
\end{array}
\end{equation}
The tensor $J_{\rho\sigma}[\Sigma]$ can also be expressed via the
4-angular momentum density tensor $J^\mu{}_{\rho\sigma}$ by the
following formula
\begin{equation}
J_{\rho\sigma}[\Sigma]=\ints J^\mu{}_{\rho\sigma}.
\end{equation}
Then
\begin{equation}
\begin{array}{l}
J^\mu{}_{\rho\sigma}=\intzi'Dz^\mu\left\{z_\rho(mDz_\sigma+\pi_\sigma)-z_\sigma(mDz_\rho+\pi_\rho)\right\}\xz\\
\qquad-\left\{x_\rho\left(\frac{\partial(\Lag^F+\Lag^I)}{\partial\phi,_\mu}\phi_\sigma-
\Lag^F\delta^\mu{}_\sigma\right)-x_\sigma\left(\frac{\partial(\Lag^F+\Lag^I)}{\partial\phi,_\mu}\phi_\rho-
\Lag^F\delta^\mu{}_\rho\right)\right\}-
\frac{i}{\hbar}\frac{\partial(\Lag^F+\Lag^I)}{\partial\phi,_\mu}S_{\rho\sigma}\phi\\
\qquad=x_\rho\left[\intzi'Dz^\mu(mDz_\sigma+\pi_\sigma)-\left(\frac{\partial(\Lag^F+\Lag^I)}{\partial\phi,_\mu}\phi_\sigma-
\Lag^F\delta^\mu{}_\sigma\right)\right]\\
\qquad-x_\sigma\left[\intzi'Dz^\mu(mDz_\rho+\pi_\rho)-\left(\frac{\partial(\Lag^F+\Lag^I)}{\partial\phi,_\mu}\phi_\rho-
\Lag^F\delta^\mu{}_\rho\right)\right]-
\frac{i}{\hbar}\frac{\partial(\Lag^F+\Lag^I)}{\partial\phi,_\mu}S_{\rho\sigma}\phi\\
\qquad=\left\{x_\rho T^\mu{}_\sigma-x_\sigma T^\mu{}_\rho\right\}+
\left\{-\frac{i}{\hbar}\frac{\partial(\Lag^F+\Lag^I)}{\partial\phi,_\mu}S_{\rho\sigma}\phi\right\}
\equiv M^\mu{}_{\rho\sigma}+S^\mu{}_{\rho\sigma}.
\end{array}
\end{equation}
One can also write down
\begin{equation}
J^\mu{}_{\rho\sigma}=J^P{}^\mu{}_{\rho\sigma}+J^I{}^\mu{}_{\rho\sigma}+J^F{}^\mu{}_{\rho\sigma},
\end{equation}
where
\begin{equation}
\begin{array}{l}
J^P{}^\mu{}_{\rho\sigma}=\left\{x_\rho T^P{}^\mu{}_\sigma-x_\sigma
T^P{}^\mu{}_\rho\right\}\equiv M^P{}^\mu{}_{\rho\sigma},\quad
S^P{}^\mu{}_{\rho\sigma}=0\\
J^I{}^\mu{}_{\rho\sigma}=\left\{x_\rho T^I{}^\mu{}_\sigma-x_\sigma
T^I{}^\mu{}_\rho\right\}+
\left\{-\frac{i}{\hbar}\frac{\partial\Lag^I}{\partial\phi,_\mu}S_{\rho\sigma}\phi\right\}
\equiv M^I{}^\mu{}_{\rho\sigma}+S^I{}^\mu{}_{\rho\sigma}\\
J^F{}^\mu{}_{\rho\sigma}=\left\{x_\rho T^F{}^\mu{}_\sigma-x_\sigma
T^F{}^\mu{}_\rho\right\}+
\left\{-\frac{i}{\hbar}\frac{\partial\Lag^F}{\partial\phi,_\mu}S_{\rho\sigma}\phi\right\}
\equiv M^F{}^\mu{}_{\rho\sigma}+S^F{}^\mu{}_{\rho\sigma}.
\end{array}
\end{equation}
\subsection{Symmetrical energy-momentum tensor}

The task of constructing the symmetrical EMT
$\sym{T}_{\mu\nu}=\sym{T}_{\nu\mu}$ is of certain interest. For a
pure field system this task was solved by \cite{Bel}. Let us
generalize the Belinfante method for the case of the system
"particle+field".

It is well known that the density of any preserved quantity is
defined with the accuracy to only the term having a form of
divergence. Thus, if for certain density $\rho^\mu_A$,
$\partial_\mu\rho^\mu_A=0$, where $A$ is the multi-index that may
characterize both space-time and internal properties of the
physical system, then for the density
$\tilde{\rho}^\mu_A\equiv\rho^\mu_A+\partial_\alpha
f^{\alpha\mu}_A$, where $f^{[\alpha\mu]}_A=f^{\alpha\mu}_A$, also
holds true $\partial_\mu\tilde{\rho}^\mu_A=0$. Both densities
lead also to the same integral preserved quantity
\begin{equation}
\tilde{Q}_A\equiv\ints\tilde{\rho}^\mu_A=
\ints\rho^\mu_A+\ints\partial_\alpha f^{\alpha\mu}_A=
\ints\rho^\mu_A\equiv Q_A,
\end{equation}
since, according to the Schwinger's lemma \cite{Sch}, for the
spatially closed systems
\begin{equation}
\int\limits_\Sigma d\sigma_\mu\,\partial_\alpha=
\int\limits_\Sigma d\sigma_\alpha\,\partial_\mu,
\end{equation}
and the quantity $f^{\alpha\mu}_A$ is anti-symmetrical over
$\alpha$ and $\mu$ indices.

Note now that from the momentum conservation law $\partial_\mu
J^\mu{}_{\rho\sigma}=0$ the following equality results
\begin{equation}
T_{[\rho\sigma]}=-\frac{1}{2}\partial_\alpha
S^\alpha{}_{\rho\sigma},
\end{equation}
i.e. the antisymmetrical part of EMT is presented in a form of
divergence. Substituting (2.39), (2.46) into (2.51), we find
\begin{equation}
\intzi'Dz_{[\rho}\frac{\partial L^I}{\partial Dz^{\sigma]}}\xz-
\frac{\partial(\Lag^F+\Lag^I)}{\partial\phi^{,[\rho}}\phi,_{\sigma]}=
\frac{1}{2}\frac{i}{\hbar}\partial_\alpha\left(\frac{\partial(\Lag^F+\Lag^I)}{\partial\phi,_{\alpha}}
S_{\rho\sigma}\phi\right).
\end{equation}
It also follows from (2.51) that
$T_{(\rho\sigma)}=T_{\rho\sigma}$, if
\begin{equation}
J^\mu{}_{\rho\sigma}=x_\rho T^\mu{}_\sigma-x_\sigma
T^\mu{}_\rho+\partial_\alpha F^{\alpha\mu}{}_{\rho\sigma},\quad
F^{[\alpha\mu]}{}_{\rho\sigma}=F^{\alpha\mu}{}_{[\rho\sigma]}=F^{\alpha\mu}{}_{\rho\sigma}.
\end{equation}
Therefore, one may write down
\begin{equation}
T^\mu{}_{\nu}=\sym{T}^\mu{}_{\nu}-\partial_\alpha
f^{\alpha\mu}{}_\nu,\quad
f^{[\alpha\mu]}{}_\nu=f^{\alpha\mu}{}_\nu.
\end{equation}
Then
\begin{equation}
\begin{array}{l}
J^\mu{}_{\rho\sigma}=(x_\rho\sym{T}^\mu{}_\sigma-
x_\sigma\sym{T}^\mu{}_\rho)- (x_\rho\partial_\alpha
f^{\alpha\mu}{}_\sigma-x_\sigma\partial_\alpha
f^{\alpha\mu}{}_\rho)+S^\mu{}_{\rho\sigma}\\
\qquad=\sym{J}^\mu{}_{\rho\sigma}-
\partial_\alpha(x_\rho f^{\alpha\mu}{}_\sigma-x_\sigma
f^{\alpha\mu}{}_\rho)+(f_\rho{}^\mu{}_\sigma-f_\sigma{}^\mu{}_\rho+S^\mu{}_{\rho\sigma}),
\end{array}
\end{equation}
where
\begin{equation}
\sym{J}^\mu{}_{\rho\sigma}\equiv x_\rho\sym{T}^\mu{}_\sigma-
x_\sigma\sym{T}^\mu{}_\rho.
\end{equation}
To satisfy (2.53) we shall require the following equation
\begin{equation}
f_\rho{}^\mu{}_\sigma-f_\sigma{}^\mu{}_\rho+S^\mu{}_{\rho\sigma}=0,
\end{equation}
or
\begin{equation}
f_{\rho\mu\sigma}-f_{\sigma\mu\rho}=-S_{\mu\rho\sigma}.
\end{equation}
to be valid. Hence,
\begin{equation}
f_{\alpha\mu\nu}=\frac{1}{2}[(f_{\alpha\mu\nu}-f_{\nu\mu\alpha})-
(f_{\mu\nu\alpha}-f_{\alpha\nu\mu})+(f_{\nu\alpha\mu}-f_{\mu\alpha\nu})]=
-\frac{1}{2}(S_{\mu\alpha\nu}+S_{\nu\alpha\mu}-S_{\alpha\mu\nu}),
\end{equation}
or, recalling formula (2.46)
\begin{equation}
f^\alpha{}_{\mu\nu}=\frac{1}{2}\frac{i}{\hbar}
\left(\frac{\partial(\Lag^F+\Lag^I)}{\partial\phi^{,\mu}}S^\alpha{}_\nu\phi+
\frac{\partial(\Lag^F+\Lag^I)}{\partial\phi^{,\nu}}S^\alpha{}_\mu\phi-
\frac{\partial(\Lag^F+\Lag^I)}{\partial\phi,_\alpha}S_{\mu\nu}\phi\right).
\end{equation}
Using formulae (2.54), (2.59), (2.60) and (2.52), we construct
$\sym{T}_{\mu\nu}$ in the manifest form
\begin{equation}
\begin{array}{l}
\sym{T}_{\mu\nu}=T_{\mu\nu}+\partial_\alpha f^\alpha{}_{\mu\nu}=
\intzi'Dz_\mu(mDz_\nu+\pi_\nu)\xz-
\left(\frac{\partial(\Lag^F+\Lag^I)}{\partial\phi^{,\mu}}\phi,_\nu-
\eta_{\mu\nu}\Lag^F\right)\\
\qquad+\frac{1}{2}\frac{i}{\hbar}\partial_\alpha\left(
\frac{\partial(\Lag^F+\Lag^I)}{\partial\phi^{,\mu}}S^\alpha{}_\nu\phi+
\frac{\partial(\Lag^F+\Lag^I)}{\partial\phi^{,\nu}}S^\alpha{}_\mu\phi\right)-
\frac{1}{2}\frac{i}{\hbar}\partial_\alpha\left(
\frac{\partial(\Lag^F+\Lag^I)}{\partial\phi,_\alpha}S_{\mu\nu}\phi\right)\\
\qquad=\intzi'MDz_\mu Dz_\nu\xz+\intzi'Dz_{(\mu}\frac{\partial
L^I}{\partial Dz^{\nu)}}\xz-
\frac{\partial(\Lag^F+\Lag^I)}{\partial\phi^{,(\mu}}\phi,_{\nu)}+\eta_{\mu\nu}\Lag^F\\
\qquad+\frac{i}{\hbar}\partial_\alpha\left(\frac{\partial(\Lag
L^F+\Lag^I)}{\partial\phi^{,(\mu}}S^\alpha{}_{\nu)}\phi\right)=T_{(\mu\nu)}-
\partial_\alpha S_{(\mu}{}^\alpha{}_{\nu)}.
\end{array}
\end{equation}
Formula (2.61) makes obvious symmetry of $\sym{T}_{\mu\nu}$.
\subsection{Invariant Hamiltonian}

Even if one fixes gauge by any Poincar\'{e}-invariant way (e.g.
imposing the condition $z'_\alpha z'^\alpha=-1$), then the action
functional (2.2) would be invariant with respect to the choice of
an initial value of the parameter, i.e. with respect to the
transformations
\begin{equation}
\left\{\begin{array}{l}
\theta\rightarrow\tilde{\theta}=\theta+\delta\varepsilon,
\quad\delta\varepsilon=\mathrm{const}\\
\bar{\delta}z^\mu(\theta)=0\\
\delta x^\mu=0\\
\delta\phi=0.
\end{array}\right.
\end{equation}
then (2.34) gives an expression for the preserved invariant
Hamiltonian of the particle (which is the generator of time
parameter translations):
\begin{equation}
\bar{\delta}W=\left.\left[\Ham[\Sigma]\right]\right|\limz\triangle\varepsilon.
\end{equation}
However, as noted above, variation of the time parameter
$\Delta\theta$ is not included into (2.34). Therefore,
substituting (2.62) into (2.34) and comparing the result with
(2.63), we obtain
\begin{equation}
\Ham=0.
\end{equation}
Thus, arbitrariness in the choice of an initial value of the time
parameter $\theta$ does not lead to the integral of motion
(additional to ten relativistic ones).

Equality to zero of the invariant Hamiltonian also indicates
impossibility to construct manifestly Poincar\'{e}-invariant and
RI-Hamiltonian formalism for the relativistic system
"particle+field", similarly to the relativistic particle that
moves in the external field. This difficulty is similar to the
"zero-Hamiltonian problem"\,in the generally covariant canonical
formulation of the general relativity \cite{Dew}. It arises in the
theory of strings \cite{Kak}, too.

Present paper was carried out within the framework of the
Ukrainian SFFR grant N $\Phi7/458-2001$.
\bibliographystyle{osa}
\bibliography{Lompay01}

\begin{thebibliography}{1}
\newcommand{\enquote}[1]{``#1''}
\expandafter\ifx\csname url\endcsname\relax
  \def\url#1{{#1}}\fi
\expandafter\ifx\csname urlprefix\endcsname\relax\def\urlprefix{}\fi

\bibitem{Kal}
G.~Kalman, Phys. Rev. {\bf 123}, 384 (1961).

\bibitem{Bel}
F.~G. Belinfante, Physica {\bf 6}, 887 (1939).

\bibitem{Sch}
J.~Schwinger, Phys. Rev. {\bf 74}, 1439 (1948).

\bibitem{Dew}
B.~S. DeWitt, Phys. Rev. {\bf 160}, 1113 (1967).

\bibitem{Kak}
M.~Kaku, {\em Introduction to Superstrings\/} (Springer-Verlag, New-York,
  1992).

\end{thebibliography}
\end{document}